\documentclass{appolb}
\usepackage[utf8]{inputenc}
\usepackage{t1enc}

\usepackage{graphicx}
\usepackage{xcolor}
\usepackage{amsmath}
\usepackage{amsfonts}

\allowdisplaybreaks

\begin{document}
\title{Hybrid star construction with the extended linear sigma model: preliminary results%
\thanks{Presented at the EQCD 2020 Conference, 2-8 February 2020}%
}
\author{Péter Kovács$^{1,2}$, János Takátsy$^{1,2}$
\address{$^{1}$Institute for Particle and Nuclear Physics, Wigner Research Centre for Physics, 1121 Budapest, Konkoly-Thege Miklós út 29-33, Hungary\\
$^{2}$Institute of Physics, Eötvös University, 1117 Budapest, Pázmány Péter stny. 1/A, Hungary}}
\maketitle

\begin{abstract}
The interior of compact stars is usually divided into two major parts, the outer part called crust and the inner part called core. There are several possibilities for the composition of these parts. One is a hybrid star, in which the crust contains nuclear matter, while the inner core contains quark matter. Since at large baryon densities one can work with effective models, and nuclear and quark matter are usually described by different models, some unification of the two parts is needed. We show two different approaches for a composite model and some recent developments in hybrid star constructions using the extended linear sigma model for modeling the quark matter at the core.
\end{abstract}

  
\section{Introduction}
In recent years much attention have been given to compact stars -- the densest objects of our Universe (the baryon density $\rho_B\approx 10^{15}- 10^{17}$~kg/m$^3$) -- especially due to the recent and ongoing gravitational wave observations. Owing to their high density mainly the strong interaction forms the matter inside the stars. It is possible that the density reaches such high values -- going towards the center of the star -- that the deconfinement transition occurs and the core of the star contains quark matter. These kind of compact stars are called hybrid stars. There are various approaches to describe this hadron--quark transition. Here we present two possible realizations already introduced in earlier studies: statistical confinement, and hadron-quark crossover with pressure interpolation. We also introduce the extended linear sigma model (eLSM) we used to describe the quark matter phase.

\section{Statistical confinement}

The first approach was introduced in \cite{benic2015} and further discussed in \cite{marczenko2018} using a hybrid quark-meson-nucleon (QMN) model, in which the deconfinement transition is realized using an auxiliary bag field that modifies the Fermi-Dirac distributions of the fermionic degrees of freedom, hence resulting in the transition between the hadronic and quark phases.

The hybrid QMN model has a Lagrangian with contributions from nucleons, mesons, quarks and the bag field. The nucleon part is a two-flavour parity doublet model with mirror assignment, where the nucleon doublets are identified with the $N(938)$ nucleon and $N(1500)$ excited nucleon states. The details and parameterization of this model was done in \cite{zschiesche2007} by fitting to properties of normal nuclear matter. The mesonic part contains scalar-isoscalar $\sigma$ ($f_0(500)$), the pseudoscalar-isovector $\pi$ ($\pi(137)$) and the vector-isoscalar $\omega_\mu$ ($\omega(783)$) mesons in a linear sigma model type Lagrangian. In the latest version of the model the $\rho$ mesons were also included \cite{marczenko2018}. The parameters of this part are fixed using tree-level meson masses and the pion decay constant. The quark part is a two-flavour linear sigma model with the appropriate Yukawa-couplings to the $\sigma$ and $\pi$ mesons, where the coupling constant can be determined from the tree-level constituent quark mass. Finally, the contribution from the bag field $b$ only contains a potential term:
\begin{equation}
    \mathcal{L} = -V_b = \frac{\kappa_b^2}{2} b^2 - \frac{\lambda_b}{4} b^4 ,
\end{equation}
where the parameters $\kappa_b$ and $\lambda_b$ are determined using the QCD vacuum energy and the pseudocritical temperature at $\mu_B = 0$.

Without introducing the bag field setting the parameters of the model to reproduce physical properties of normal nuclear matter results in a too shallow potential for the $\sigma$ meson, which causes the chiral phase transition to take place below normal nuclear saturation density at zero temperature, which is definitely not physical. To solve this, we can introduce a so-called statistical confinement, where the Fermi-Dirac distributions are modified. At zero baryochemical potential ($\mu_B$) and finite temperature the Polyakov-loop can be used, however it does not affect the zero temperature region with finite chemical potential. Instead, here we can modify the distribution functions using the bag field in the following manner:


\begin{align}
    n_{N_{\pm}} &= \Theta(\alpha^2 b^2 - \mathbf{p}^2) f_{N_{\pm}} , \quad
    \bar{n}_{N_{\pm}} = \Theta(\alpha^2 b^2 - \mathbf{p}^2) \bar{f}_{N_{\pm}} ,\\
    n_{q} &= \Theta(\mathbf{p}^2 - b^2) f_{q} , \quad
    \bar{n}_{q} = \Theta(\mathbf{p}^2 - b^2) \bar{f}_{q} ,\\
    f_x &= \frac{1}{1+e^{\beta(E_x-\mu_x)}} , \quad
    \bar{f}_x = \frac{1}{1+e^{\beta(E_x+\mu_x)}} ,\quad x\in (N_{+}, N_{-}, q)
\end{align}

where $\alpha$ is a free parameter, $\mathbf{p}$ is the three momentum, $E_x=\sqrt{\mathbf{p}^2 + m_x^2}$, and $\mu_x$ is the chemical potential for $x\in (N_{+}, N_{-}, q)$. The effect of the modified distribution functions is the suppression of quarks at low momentum, which makes them to appear only at higher densities despite the low value of the $m_\sigma$ mass. On the other hand nucleons will be suppressed at high momentum, which subsequently results in their disappearance at higher densities.

The grand potential is calculated using a mean-field approximation and the gap equations are determined by considering $b$ as a dynamical field as well. At high densities $\partial b/\partial\mu_B < 0$, which gradually suppresses the contribution from nucleons. After determining the zero-temperature equation of state and solving the Tolman-Oppenheimer-Volkoff equations the obtained $M-R$ curves are consistent with current astrophysical measurements. As a criticism of this model one might mention that it is difficult to reveal the origin of such a bag field in fundamental theories. Moreover, while we might expect the chiral and deconfinement transitions to occur close to each other, according to lattice calculations at zero density, in this model there is a significant gap between them ($\Delta \mu_B \approx 600$ MeV).

\section{Hadron\,-\,quark crossover with $p$\,-\,interpolation}

The second method was proposed in \cite{masuda2013} and applies a smooth interpolation between the $p(\varepsilon)$ curves of the two phases. This approach assumes that the quark degrees of freedom appear gradually with increasing density, as the finite-size nucleons start to overlap. At intermediate densities neither the low density hadronic nor the high density quark equation of state is applicable, instead one might want to find an appropriate interpolation between the two regions.

In \cite{masuda2013} the authors use the TNI2u and TNI3u hadronic equations of state \cite{nishizaki2001,nishizaki2002}, which were obtained using G-matrix calculations with two- and three-body interactions, similarly to the APR equation of state, while hyperons were also included in these models. The quark equation of state is obtained from a three-flavour Nambu-Jona-Lasinio model with vector meson interactions (see {\it e.g.} \cite{klevansky1992}).

The pressure -- as a function of baryon density $\rho_B$ -- is defined in the intermediate region as,
\begin{align}
    p(\rho_B) &= p_H(\rho_B) f_-(\rho_B) + p_q(\rho_B) f_+(\rho_B) ,\\
    f_\pm(\rho_B) &= \frac{1}{2} \left( 1 \pm \tanh\left(\frac{\rho_B - \bar{\rho}_B}{\Gamma}\right) \right) ,
\end{align}
where $\bar{\rho}_B$ is the middle of the intermediate region and $\Gamma$ is its width. To have a thermodynamically consistent equation of state we need to derive the expression for the energy density from the pressure using the relation $p = \rho_B^2 \, \partial(\varepsilon/\rho_B)/\partial \rho_B$. This induces the following expression for $\varepsilon$:
\begin{align}
    \varepsilon(\rho_B) &= \varepsilon_H(\rho_B) f_-(\rho_B) + \varepsilon_q(\rho_B) f_+(\rho_B) + \Delta\varepsilon ,\\
    \Delta\varepsilon &= \rho_B \int_{\bar{\rho}_B}^{\rho_B} (\varepsilon_H(\rho') - \varepsilon_q(\rho')) \frac{g(\rho')}{\rho'} \mathrm{d}\rho' ,\\
    g(\rho') &= \frac{1}{2\Gamma} \cosh^{-2}\left(\frac{\rho' - \bar{\rho}_B}{\Gamma}\right) .
\end{align}
Using only the hadronic equation of state with hyperons, as it is well
known, one cannot obtain neutron stars with $2$ solar masses. However,
with this interpolation it is possible to get such high-mass neutron
stars while preserving causality. In addition to the highly
phenomenological nature of this approach another criticism is that the
authors of \cite{masuda2013} consider a crossover starting at
densities as low as $2\rho_0$ ($\rho_0=0.16\,\text{fm}^{-3}$), while
nuclear collisions do not show a significant overlap in the
wave-functions of nucleons at such low densities.

\section{The extended linear sigma model}

The model we use here to describe the quark sector is an $N_f = 2+1$ flavour (axial)vector meson extended linear sigma model (eLSM). The Lagrangian and the detailed description of this model, in which in addition to the full nonets of (pseudo)scalar mesons the nonets of (axial)vector mesons are also included, can be found in \cite{kovacs2016,parganlija2013}. The model contains three flavours of constituent quarks, with kinetic terms and Yukawa-type interactions with the (pseudo)scalar mesons. In addition to the model described in \cite{kovacs2016}, a Yukawa term is also included between the quark and vector meson sector, and a condensate for $\omega^0$ is also introduced, as in \cite{takatsy2019}.

The grand potential is calculated using a mean-field approximation, in which fermionic fluctuations are included at one-loop level, while the mesons are treated at tree-level. The parameters of the model are determined with a $\chi^2$ fit to tree-level meson masses and two-body decay widths, while requiring that the model reproduce the lattice value of the pseudocritical temperature at $\mu_B=0$. The quark-vector meson coupling $G_V$ is varied between 0 and 3. A higher $G_V$ value makes the equation of state stiffer, hence producing higher-mass compact stars.

\section{Results}

\begin{figure}[!t]
\centerline{%
\includegraphics[width=10.5cm]{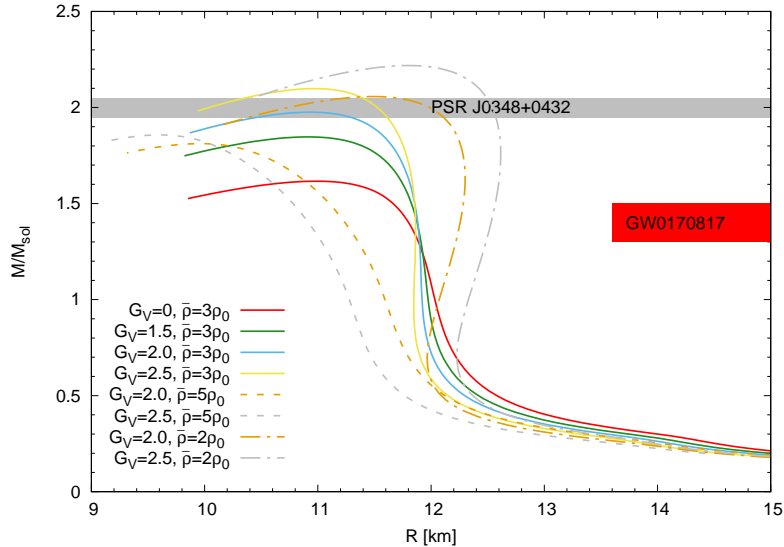}}
\caption{Mass--radius diagrams of hybrid stars constructed from the TNI3u and eLSM equations of state using the pressure interpolation method. The different curves show the diagrams obtained from using different values for $\bar{\rho}_B$ and for the $G_V$ quark-vector meson coupling.}
\label{Fig:MR_eLSM_hybrid}
\end{figure}

In our application we used the pressure interpolation method to construct hybrid star equations of state from the TNI3u model and the eLSM. We used the BPS equation of state to describe the low-density crust. We used different values for the quark-vector meson coupling $G_V$ and for $\bar{\rho}_B$, while we fixed the value of $\Gamma = \rho_0$. Then we solved the Tolman-Oppenheimer-Volkoff equations to obtain the mass--radius relations. These diagrams are shown in Fig. \ref{Fig:MR_eLSM_hybrid} together with the 2 solar mass constraint and the radius constraint from GW170817 \cite{PhysRevLett.120.172703}. We see that for high-enough values of the vector meson coupling the curves reach the 2 solar mass limit, while they have an almost vertical region (with $R\approx$ const.) for $\bar{\rho}_B=3\rho_0$.

\section{Conclusion}

We gave an overview of several methods of how one can construct hybrid star equations of state from a low-density hadronic and a high-density quark model. We also presented various mass--radius curves using the pressure interpolation method to attach the equations of state from the TNI3u model and the eLSM. We see that current astrophysical constraints can be satisfied in many ways and further, more precise radius measurements would be required to narrow down these possibilities. It is worth to note that the quark-vector meson coupling has a significant effect on the stiffness of the equation of state, although even pure quark stars are not completely excluded currently \cite{takatsy2019}. 

\section*{Acknowledgement}

P. Kovács acknowledges support by the János Bolyai Research Scholarship of the Hungarian Academy of Sciences and was also supported by the ÚNKP-19-1 New National Excellence Program of the Ministry for Innovation and Technology. J. Takátsy acknowledges support by the NRDI fund of Hungary, financed under the FK\textunderscore19 funding scheme, project no. FK 131982.

\bibliographystyle{unsrt}
\bibliography{EQCD}

\end{document}